\documentstyle[12pt]{article}

\input{psfig}
\renewcommand{\baselinestretch}{1.5}
\renewenvironment{thebibliography}[1]
{\small
\begin{list}{[\arabic{enumi}]}
{\usecounter{enumi}\setlength{\parsep}{0pt}
\setlength{\leftmargin 1cm}{\rightmargin 1cm}
\setlength{\itemsep}{0.1cm}
\settowidth{\labelwidth}{#1.}\sloppy}}{\end{list}}
\newcommand{\lesssim}
{\lower 2pt \hbox{$\:\stackrel{<}{\scriptstyle \sim}\:$}}
\newcommand{\gtrsim}
{\lower 2pt \hbox{$\:\stackrel{>}{\scriptstyle \sim}\:$}}

\begin{document}

\renewcommand{\baselinestretch}{0.6}
\begin{scriptsize}
\noindent
\parbox{12cm}{
To appear in the proceedings of the XXIXth RENCONTRES DE MORIOND on
{\em Coulomb and Interference Effects in Small Electronic Structures},
January 1994, Villars-sur-Ollon, Switzerland,
edited by D. C. Glattli and M. Sanquer
(Editions Fronti\`{e}res, France).
\mbox{\tt cond-mat/9404040}}
\end{scriptsize}
\bigskip
\bigskip

\begin{center}
{\bf SUB-POISSONIAN SHOT NOISE IN A DIFFUSIVE CONDUCTOR}

\bigskip
\medskip

{\bf M. J. M. de Jong$^{a,b}$ and C. W. J. Beenakker$^{b}$}
\\
{\em
(a) Philips Research Laboratories,
5656 AA  Eindhoven,
The Netherlands\\
(b) Instituut-Lorentz,
University of Leiden,
2300 RA  Leiden,
The Netherlands
}

\bigskip
\bigskip

\renewcommand{\baselinestretch}{1.2}
\begin{small}
\parbox{15cm}{
{\bf Abstract.}\ --- A review is given of the shot-noise properties of
metallic, diffusive conductors.
The shot noise is one third of the Poisson noise, due to
the bimodal distribution of transmission eigenvalues.
The same result can be obtained from a semiclassical calculation.
Starting from Oseledec's theorem it is shown that the bimodal distribution is
required by Ohm's law.}
\end{small}
\end{center}

\bigskip
\noindent
{\bf I. Introduction}
\medskip

Time-dependent
fluctuations in the electrical current caused
by the discreteness of the charge carriers are known as shot noise.
These fluctuations are characterized by
a white noise spectrum and persist down to zero temperature.
The noise spectral density $P$ (per unit frequency bandwidth)
is a measure for the magnitude of these fluctuations.
A well-known example is a saturated vacuum diode, for which Schottky
found that
$P=2 e I\equiv P_{\rm Poisson}$, with $I$ the average current \cite{sch18}.
This indicates
that the electrons traverse the conductor as uncorrelated
current pulses,
i.e.\ are transmitted in time according to Poisson statistics.
It is also known that a metal wire, of macroscopic
length $L$,
does not exhibit shot noise, because inelastic scattering
reduces $P$ by a factor $l_i/L$, which is much smaller than 1
in a macroscopic conductor ($l_i$ is the
inelastic scattering length).
In the last decade, the investigation of transport on smaller length scales
has become accessible through
the progress in microfabrication techniques.
The physics on this mesoscopic scale displays a wealth of new
phenomena \cite{c&h,MPiS}.
Theoretical analysis \cite{khl87,les89,yur90,but90}
shows that the shot noise
in mesoscopic conductors may be suppressed below $P_{\rm Poisson}$,
due to correlated electron transmission as a consequence of the
Pauli principle.
This raises the question how large $P$ is
in a {\em metallic, diffusive} conductor of length $L < l_i$,
but still longer than the elastic mean free
path $\ell$.
It has been predicted theoretically \cite{b&b92,nag92,jon92}
that $P= \frac{1}{3} P_{\rm Poisson}$.
This suppression of the shot noise
by a factor one third is {\em universal},
in the sense that it does not depend on the specific geometry nor
on any intrinsic material parameter (such as $\ell$).
The purpose of this paper is to discuss the
origin of the one-third suppression.
First, we review the fully quantum-mechanical
calculation, where
the suppression originates from the
bimodal distribution of transmission eigenvalues.
Then, a semiclassical calculation is presented, which surprisingly yields
the same suppression by one third.
One might therefore ask whether there exists a semiclassical explanation
for the bimodal eigenvalue distribution.
Indeed, we find that this distribution
is required by Ohm's law.
We conclude with a brief discussion of an
experimental observation of suppressed shot noise in a disordered wire,
which has recently been reported \cite{lie94}.

\bigskip
\noindent
{\bf II. Quantum-mechanical theory}
\medskip

A scattering formula for the shot noise in a phase-coherent conductor has been
derived by B\"{u}ttiker \cite{but90}.
It relates the
zero-temperature, zero-frequency shot-noise
power $P$
of a spin-degenerate, two-probe conductor
to the transmission matrix ${\sf t}$:
\begin{equation}
P = P_0 \mbox{Tr} \,
[ {\sf t}  {\sf t}^\dagger
( {\sf 1} - {\sf t}
{\sf t}^\dagger ) ]
= P_0 \sum_{n=1}^{N}
T_n ( 1 - T_n ) \; .
\label{e1}
\end{equation}
Here $P_0 \equiv 2 e V  (2 e^2 / h)$, with $V$ the applied voltage,
$T_n$ denotes an eigenvalue
of ${\sf t} {\sf t}^\dagger$, and
$N$ is the number of transverse modes at the Fermi energy $E_F$.
It follows from current conservation that the
transmission eigenvalues $T_n \in [0,1]$.
Equation (\ref{e1}) is the multi-channel generalization of
single-channel formulas found earlier \cite{khl87,les89,yur90}.
Levitov and Lesovik have shown \cite{lev93} that
Eq.\ (\ref{e1}) follows from the fact that the
electrons in each separate scattering channel
are transmitted in time according to a
binomial (Bernoulli) distribution (depending on $T_n$).
The Poisson noise is then just a result of the limiting distribution
for small $T_n$.
Using the Landauer formula for the conductance
\begin{equation}
G = G_0
\mbox{Tr}\, {\sf t} {\sf t}^\dagger
 = G_0 \sum_{n=1}^{N} T_n \; ,
\label{e2}
\end{equation}
with $G_0 \equiv 2 e^2/h$, one finds
from Eq.\ (\ref{e1}) that indeed $P = 2 e V G = 2 e I = P_{\rm Poisson}$
if $T_n \ll 1$ for all $n$.
However, if the transmission eigenvalues are not much smaller than 1,
the shot noise is suppressed below $P_{\rm Poisson}$.
As mentioned above, this suppression is a consequence of the electrons
being fermions. In a scattering channel with $T_n \ll 1$ the electrons are
transmitted in time
in uncorrelated fashion. As $T_n$ increases the electron
transmission becomes more correlated because of the Pauli principle.
In a scattering channel with $T_n=1$ a constant current is flowing, so that
its contribution to the shot noise is zero.

Let us now turn to transport through a diffusive conductor
($L \gg \ell$), in the metallic regime ($L \ll$ localization length).
To compute the ensemble averages $\langle \cdots \rangle$
of Eqs.\ (\ref{e1}) and (\ref{e2}) we need the density of
transmission eigenvalues
$p(T) = \langle \sum_n \delta(T-T_n) \rangle$.
The first moment of $p(T)$ determines the conductance,
\begin{equation}
\langle G \rangle =  G_0 \int \limits_0^1 dT \, p(T) \, T \: ,
\label{e2c}
\end{equation}
whereas the shot-noise power contains also the second moment
\begin{equation}
\langle P \rangle =  P_0 \int \limits_0^1 dT \, p(T) \, T(1-T) \: .
\label{e2d}
\end{equation}
In the metallic regime, Ohm's law for the conductance holds to a good
approximation,
which implies that $\langle G \rangle \propto 1/L$, up to small corrections
of order $e^2/h$ (due to weak localization).
The Drude formula gives
\begin{equation}
\langle G \rangle=G_0 \, \frac{N \tilde{\ell}}{L} \: ,
\label{e3}
\end{equation}
where $\tilde{\ell}$ equals the mean free path $\ell$ times
a numerical coefficient \cite{note}.
 From Eqs.\ (\ref{e2c}) and (\ref{e3}) one might surmise that
for a diffusive conductor all the transmission eigenvalues are
of order $\tilde{\ell} /L$, and hence much smaller than 1.
This would imply the shot-noise power $P=P_{\rm Poisson}$ of
a Poisson process.

\begin{figure}
%
%
\centerline{\psfig{file=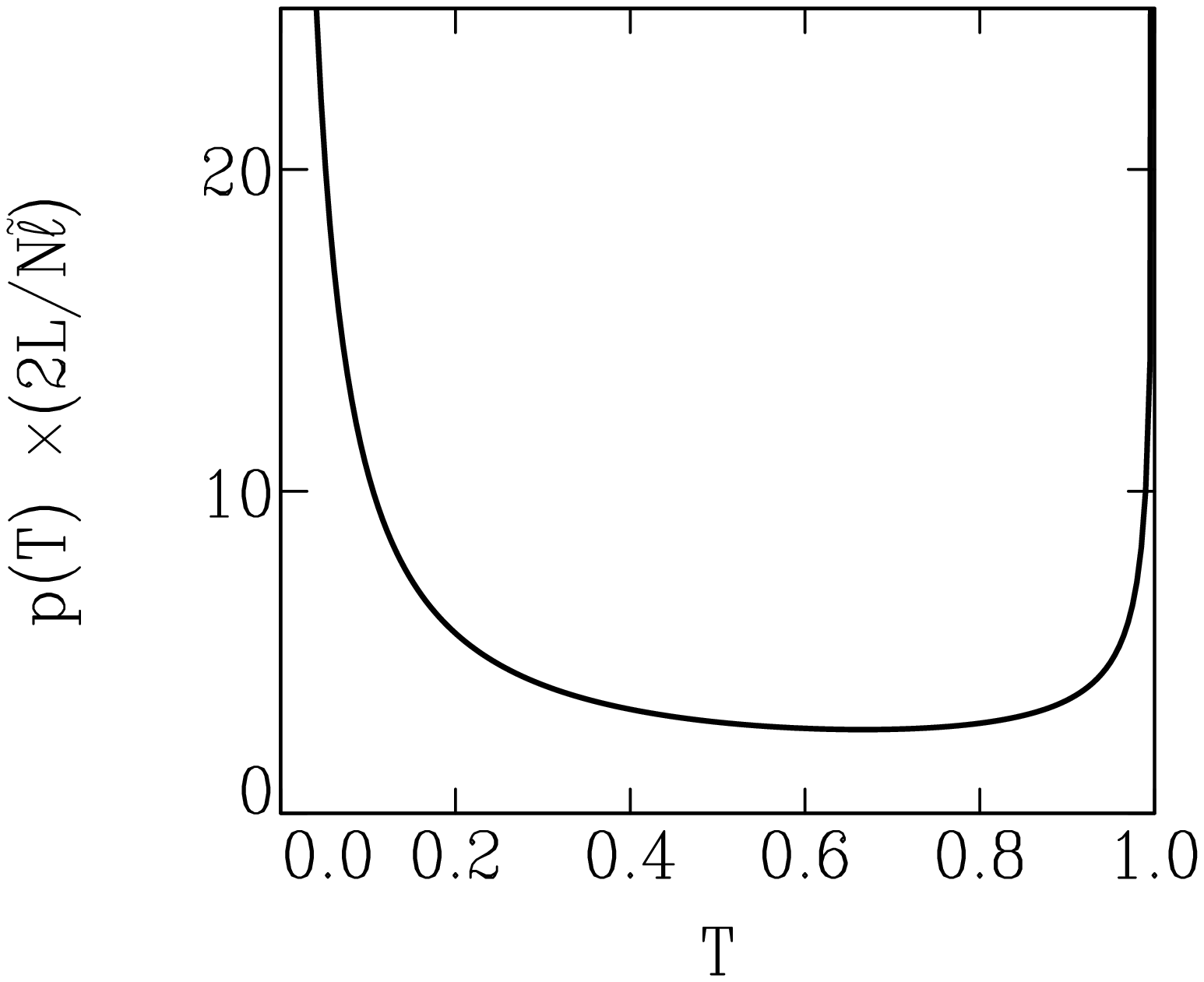,width=10cm}}
\vspace{0.2cm}
\refstepcounter{figure}
\begin{center}
\renewcommand{\baselinestretch}{1.2}
\begin{small}
\parbox{15cm}{
{\bf Figure~\thefigure.}
The bimodal distribution of transmission eigenvalues
according to Eq. (\protect{\ref{e5}}).
The cutoff for $T \lesssim 4 \exp(-2L/\tilde{\ell})$ is not shown.
}
\end{small}
\end{center}
\label{f1}
\end{figure}

However,
the surmise $T_n \approx \tilde{\ell} /L$ for all $n$ is completely
incorrect for a metallic, diffusive conductor.
This was first pointed out by Dorokhov \cite{dor84},
and later by Imry \cite{imr86} and by Pendry {\em et al.} \cite{pen92}.
In reality,  a fraction
$\tilde{\ell}/L$ of the transmission eigenvalues is of order unity
(open channels),
the others being exponentially small (closed channels).
The full distribution function is
\begin{equation}
p(T) = \frac{N \tilde{\ell}}{2 L} \, \frac{1}{T \sqrt{1-T}} \,
\Theta( T - T_0 ) \: ,
\label{e5}
\end{equation}
where $T_0 \simeq 4 \exp(- 2 L/\tilde{\ell}) \ll 1$
is a cutoff at small $T$ such that $\int_0^1 dT \, p(T) = N$
(the function $\Theta(x)$ is the unit step function).
One easily checks that Eq.\ (\ref{e5}) leads to the Drude conductance
(\ref{e3}).
The function $p(T)$ is plotted in Fig.\ \ref{f1}.
It is {\em bimodal} with peaks near unit and zero transmission.
The distribution (\ref{e5})
follows from a scaling
equation, which describes the evolution
of $p(T)$ on increasing $L$ \cite{dor82,mel88a,stoMP}.
A microscopic derivation of Eq.\ (\ref{e5}) has recently been given
by Nazarov \cite{naz94}.

The bimodal distribution (\ref{e5}) implies for the shot-noise power
(\ref{e2d})
the unexpected result \cite{b&b92}
\begin{equation}
\langle P \rangle=
 \frac{1}{3} P_0 \, \frac{N \tilde{\ell}}{L} = \frac{1}{3} P_{\rm Poisson}
\; .
\label{e6}
\end{equation}
Corrections to Eq.\ (\ref{e6}) due to weak localization have also been
computed \cite{jon92}, and are smaller by a factor $L/N \tilde{\ell}$
(which is $\ll 1$ in the metallic regime).

\bigskip
\noindent
{\bf III. Semiclassical calculation}
\nopagebreak
\medskip

Since the Drude conductance (\ref{e3}) can be obtained semiclassically
(without taking quantum-interference effects into account),
one may wonder whether the sub-Poissonian shot noise (\ref{e6})
--- which follows from the same $p(T)$ --- might also be obtained from a
semiclassical calculation.
Such a calculation was presented by
Nagaev \cite{nag92}, who independently from
Refs.\ \cite{b&b92,jon92} arrived at the result (\ref{e6}).
Nagaev uses a Boltzmann-Langevin approach \cite{kad57,kog69}, which is a
classical kinetic theory for the non-equilibrium
fluctuations in a degenerate electron gas. We refer to
this method as semiclassical, because the motion of the electrons
is treated classically ---
without quantum-interference effects ---
whereas the Pauli principle is accounted for,
through the use of Fermi-Dirac statistics.
Nagaev's approach does not yield a formula with the same generality as
B\"{u}ttiker's formula (\ref{e1}),
but is only applicable for diffusive transport.

\begin{figure}[tb]
%
%
\centerline{\psfig{file=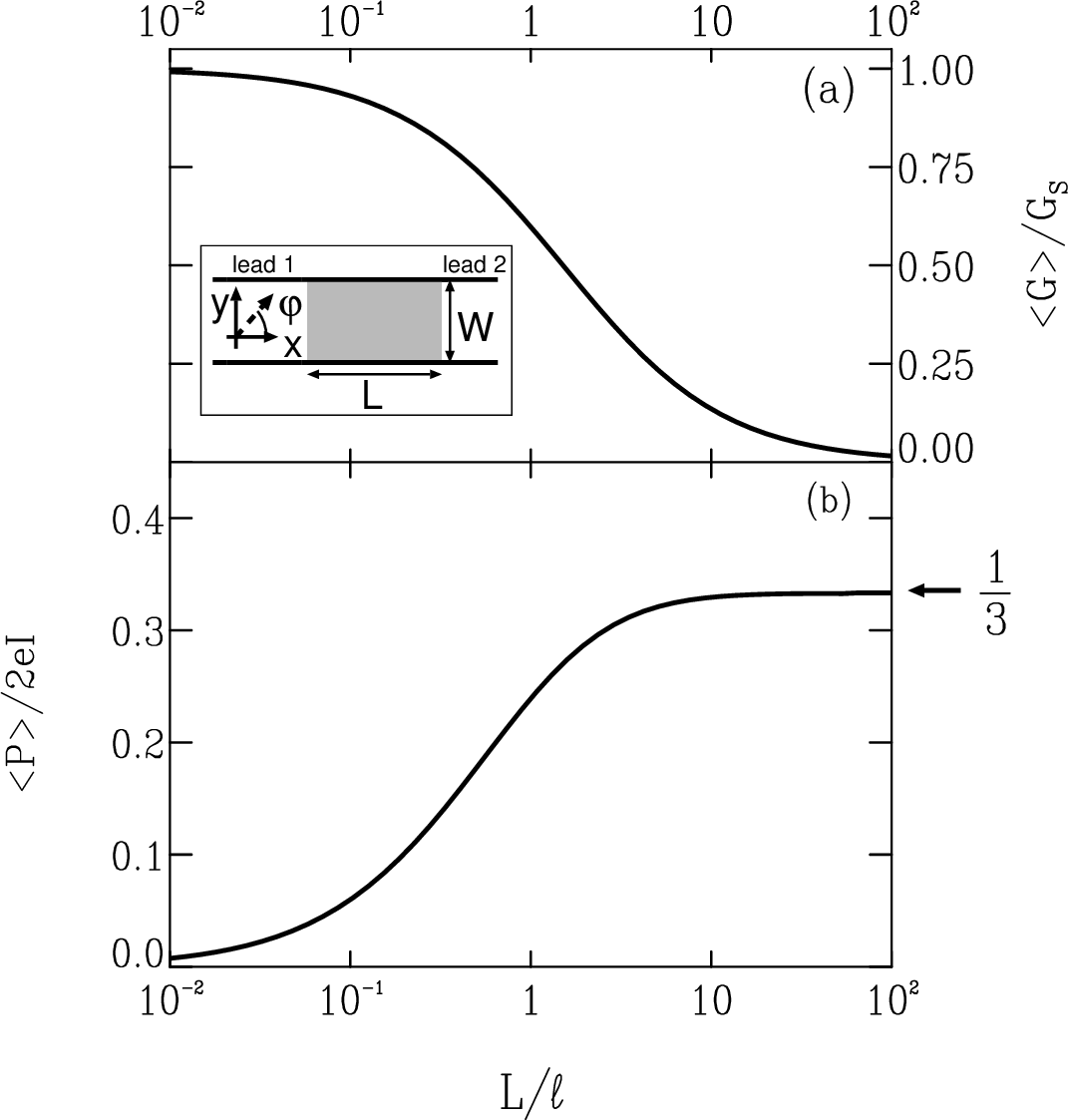,width=12cm}}
\vspace{0.2cm}
\refstepcounter{figure}
\begin{center}
\renewcommand{\baselinestretch}{1.2}
\begin{small}
\parbox{15cm}{
{\bf Figure~\thefigure.}
(a) The conductance (normalized by the Sharvin conductance
$G_S \equiv N G_0$) and (b) the shot-noise power
(in units of $P_{\rm Poisson} \equiv 2 e I$),
as a function of the ratio $L/\ell$,
computed from Eqs.\ (\ref{e7}) and (\ref{e8})
for isotropic impurity scattering.
The inset shows schematically the wire and its coordinates.
}
\end{small}
\end{center}
\label{f2}
\end{figure}

To put the quantum-mechanical and the semiclassical theories of shot noise
on equal terms,
we have recently derived a scattering formula for $P$
from the Boltzmann-Langevin approach.
This formula is valid from the ballistic to the
diffusive transport regime.
A detailed description will be the subject of a forthcoming publication.
Here, we merely present the result.
For simplicity, we consider a two-dimensional wire
(length $L$ and width $W$),
with a circular Fermi surface.
The geometry is shown in Fig.\ \ref{f2} (inset).
The scattering formula relates $P$ to the classical
transmission probabilities $T({\bf r}, \varphi)$, which denote the probability
that an electron at position ${\bf r} \equiv (x,y)$ with velocity
${\bf v} \equiv v_F(\cos \varphi, \sin \varphi)$
(with $v_F$ the Fermi velocity) is transmitted
into lead number 2.
The result is
\begin{equation}
P= \frac{N P_0}{4 \pi W v_F} \,
\int \limits_0^L dx \int \limits_0^W dy
\int \limits_0^{2 \pi} d \varphi
\int \limits_0^{2 \pi} d \varphi' \,
W_{\varphi \varphi'}({\bf r}) \,
[ T({\bf r},\varphi) - T({\bf r},\varphi') ]^2 \,
\overline{T}({\bf r},\varphi) [ 1 - \overline{T}({\bf r},\varphi') ] \; ,
\label{e7}
\end{equation}
where the number of channels $N=  W m v_F/ \hbar \pi$,
and $W_{\varphi \varphi'}({\bf r})$ is the transition rate for
(elastic) impurity-scattering from
$\varphi$ to $\varphi'$, which may in principle
depend also on ${\bf r}$.
The time-reversed probability $\overline{T}({\bf r},\varphi)$
gives the probability
that an electron at $({\bf r},\varphi)$ has originated from \mbox{lead 2}.
{}From now on we assume time-reversal symmetry (zero magnetic field), so that
$\overline{T}({\bf r},\varphi)=T({\bf r},\varphi+\pi)$.
Equation (\ref{e7}) corrects a previous result \cite{bee91}.
In this notation,
the conductance is given by
\begin{equation}
G= \frac{N G_0}{2 W}  \int \limits_0^W dy
\int \limits_0^{2 \pi} d \varphi \, \cos \varphi \: T({\bf r}, \varphi) \; .
\label{e8}
\end{equation}
Eq.\ (\ref{e8}) is independent of $x$ because of current conservation.
The transmission probabilities obey a Boltzmann type of
equation \cite{jon94}
\begin{equation}
{\bf v} \cdot
\mbox{\boldmath $\nabla$}T({\bf r}, \varphi) = \int \limits_0^{2 \pi}
\frac{d \varphi'}{2 \pi} \, W_{\varphi \varphi'}({\bf r}) \,
\left[ T({\bf r}, \varphi) - T({\bf r}, \varphi') \right] \: ,
\label{e8a}
\end{equation}
where $\mbox{\boldmath $\nabla$} \equiv (\partial/\partial x,
\partial/\partial y)$.

We now apply Eq.\ (\ref{e7}) to the case
$W_{\varphi \varphi'}({\bf r})=v_F/\ell$
of isotropic impurity scattering.
Since the scattering is
modeled by one parameter, the resulting $P$ is the ensemble average.
We assume specular boundary scattering, so that the transverse
coordinate ($y$) becomes irrelevant.
Let us first show that in the
diffusive limit ($\ell \ll L$) the result of Nagaev \cite{nag92} is recovered.
For a diffusive wire the solution of Eq.\ (\ref{e8a})
 can be approximated by
\begin{equation}
T({\bf r},\varphi) = \frac{x + \ell \cos \varphi}{L} \; .
\label{e9}
\end{equation}
Substitution into Eq.\ (\ref{e8}) yields the Drude
conductance $\langle G \rangle = N G_0 \, \pi \ell / 2 L$
in accordance with Eq.\ (\ref{e3}).
For the shot-noise power one obtains, neglecting terms of order
$(\ell/L)^2$,
\begin{equation}
\langle P \rangle = N P_0 \, \frac{\pi \ell}{L} \int \limits_0^L \frac{dx}{L}
\,
\frac{x}{L} \left( 1 - \frac{x}{L} \right)
= \frac{1}{3} P_{\rm Poisson}
\; ,
\label{e10}
\end{equation}
in agreement with Eq.\ (\ref{e6}).

We can go beyond Ref.\ \cite{nag92} and apply our method to
quasi-ballistic wires, for which $\ell$ and $L$ become comparable.
In Ref.\ \cite{jon94} it is shown how in this case the
probabilities $T({\bf r}, \varphi)$
can be calculated numerically by solving Eq.\ (\ref{e8a})
through Milne's equation.
In Fig.\ \ref{f2} we show the result for both the conductance and the
shot-noise power.
The conductance crosses over
from the Sharvin conductance ($G_S \equiv N G_0$) to the Drude conductance
with increasing wire length \cite{jon94}.
This crossover is accompanied by a rise in the shot noise,
from zero to $\frac{1}{3} P_{\rm Poisson}$.

\bigskip
\noindent
{\bf IV. Bimodal eigenvalue distribution from Ohm's law}
\medskip

Now that it is established that the
quantum-mechanical calculation (Sec.\ II) and the semiclassical approach
(Sec.\ III) yield the one-third suppression of the shot noise,
we would like to close the circle by showing how the bimodal distribution
(\ref{e5}) of the transmission eigenvalues can be obtained
semiclassically.

It is convenient to work with the parametrization
\begin{equation}
T_n = \frac{1}{\cosh^2 (\alpha_n L)} \: , \;\;\;\; n=1,2, \ldots N \: ,
\label{e11}
\end{equation}
which relates the eigenvalues $T_n$ of ${\sf t t}^\dagger$ to the
eigenvalues $\exp( \pm 2 \alpha_n L )$ of ${\sf M M}^\dagger$.
Here $\sf t$ is the $N \times N$ transmission matrix, $\sf M$ is the
$2N \times 2N$ transfer matrix of the conductor,
and $\alpha_n \in [0, \infty)$ for all $n$.
The eigenvalues of ${\sf M M}^\dagger$ come in inverse pairs as a result of
current conservation \cite{stoMP}.
The $\alpha_n$'s are known as the inverse localization lengths of
the conductor. Scattering channels for which the localization length is
longer than the sample length ($\alpha_n L \ll 1$) are open, if the
sample length exceeds the
localization length ($\alpha_n L \gg 1$) the scattering channel
is closed, as is clear from Eq.\ (\ref{e11}).
The bimodal distribution (\ref{e5}) of the transmission eigenvalues
is equivalent to a
{\em uniform} distribution of the inverse localization lengths,
\begin{equation}
\rho(\alpha) = N \tilde{\ell} \, \Theta(\alpha - 1/\tilde{\ell}) \: ,
\label{e12}
\end{equation}
where $\rho(\alpha) \equiv \langle \sum_n \delta(\alpha-\alpha_n) \rangle$.
Furthermore, the distribution of the $\alpha$'s implied
by Eq.\ (\ref{e12}) is {\em independent} of the sample length $L$.
We will argue that these two properties, $L$-independence and
uniformity, of $\rho(\alpha)$
follow from Oseledec's theorem \cite{osl68} and Ohm's law,
respectively.

We recall \cite{stoMP} that
the transfer matrix has the multiplicative property that
if two pieces of wire with matrices ${\sf M}_1$ and
${\sf M}_2$ are connected in series,
the transfer matrix of the combined system is simply the
product  ${\sf M}_1 {\sf M}_2$.
In this way the transfer matrix of a disordered wire can be
constructed from the product of $N_L$ individual transfer
matrices ${\sf m}_i$,
\begin{equation}
{\sf M} = \prod_{i=1}^{N_L} {\sf m}_i \: ,
\label{e13}
\end{equation}
where $N_L\equiv L/\lambda$ is a large number proportional to $L$.
The ${\sf m}_i$'s are assumed to be independently and identically
distributed random matrices, each representing transport through a
slice of conductor of small, but still macroscopic, length $\lambda$.
In the theory of random matrix products \cite{ProRM}, the limits
$\lim_{L \rightarrow \infty} \alpha_n$ are known as the Lyapunov exponents.
Oseledec's theorem \cite{osl68} is the statement that this limit exists.
Numerical simulations \cite{stoMP} indicate that the large-$L$ limit is
essentially reached for $L \gg \ell$, and does not require
$L \gg N \ell$. This explains the $L$-independence of the distribution of
the inverse localization lengths
 in the metallic, diffusive regime ($\ell \ll L \ll N \ell$).

Oseledec's theorem tells us that $\rho(\alpha)$ is independent of $L$,
but it does not tell us how it depends on $\alpha$.
To deduce the uniformity of $\rho(\alpha)$ we invoke Ohm's law,
$\langle G \rangle \propto 1/L$. This requires
\begin{equation}
L \int \limits_0^\infty d\alpha \, \rho(\alpha) \,
\frac{1}{\cosh^2 (\alpha L)} = C \; ,
\label{e14}
\end{equation}
where $C$ is independent of $L$\@. It is clear that
Eq.\ (\ref{e14}) implies the uniform distribution $\rho(\alpha)=C$.
A cutoff at large $\alpha$ is allowed, since $1/\cosh^2(\alpha L)$
vanishes anyway for $\alpha L \gg 1$. From Drude's formula
(\ref{e2d}) we deduce $C=N \tilde{\ell}$, and normalization then implies a
cutoff at $\alpha \gtrsim 1/\tilde{\ell}$, in accordance with
Eq.\ (\ref{e12}).

\bigskip
\noindent
{\bf V. Conclusion}
\medskip

In summary, we have discussed the equivalence of the fully
quantum-mechanical and the semiclassical theories of sub-Poissonian
shot noise in a metallic, diffusive conductor.
Both approaches yield a one-third suppression of $P$ relative to
$P_{\rm Poisson}$.
The bimodal distribution, which is at the heart of the quantum-mechanical
explanation, can be understood semiclassically as a consequence of
a mathematical theorem on eigenvalues (Oseledec) and a law of
classical physics (Ohm's law).

The fact that phase coherence is not essential for the one-third
suppression of $P$ suggests that this phenomenon is more robust than other
mesoscopic phenomena, such as universal conductance fluctuations.
This might explain the success of the recent attempt
to measure the shot-noise suppression due to open scattering channels in a
disordered wire defined in a 2D electron gas \cite{lie94}.
In this experiment a rather large current was necessary to obtain a
measurable shot noise, and it seems unlikely that phase coherence
was maintained under such conditions.

In both the quantum-mechanical and semiclassical theories discussed in
this review, the effects of electron-electron interactions have been ignored.
The Coulomb repulsion is known to have a strong effect on the noise in
confined geometries with a small capacitance \cite{her93}.
We would expect the interaction effects to be less important in open
conductors \cite{but93}.
While a fully quantum-mechanical theory of shot noise with
electron-electron interactions
seems difficult, the semiclassical Boltzmann-Langevin approach discussed
here might well be extended to include electron-electron scattering
and screening effects.

\bigskip
\noindent
{\bf Acknowledgements}
\nopagebreak
\medskip

The authors would like to thank H. van Houten, R. Landauer, and L. W.
Molenkamp for valuable discussions.
This research
was supported by the ``Ne\-der\-land\-se
or\-ga\-ni\-sa\-tie voor We\-ten\-schap\-pe\-lijk On\-der\-zoek'' (NWO)
and by the ``Stich\-ting voor Fun\-da\-men\-teel On\-der\-zoek der
Ma\-te\-rie'' (FOM).

\bigskip
\noindent
{\bf References}

\renewcommand{\baselinestretch}{1.2}


\begin{thebibliography}{00}
\bibitem{note}
The precise relationship between $\ell$ and $\tilde{\ell}$
 depends on the dimensionality of the Fermi surface and on
the specific kind of scattering.
For isotropic impurity scattering with mean free path $\ell$ one has
$\tilde{\ell} = \pi \ell/2$ in 2D (Fermi circle)
and $\tilde{\ell}=4\ell/3$ in 3D (Fermi sphere).
\end{thebibliography}
\end{document}